\documentclass[referee]{mn2e}
\usepackage{graphics}
\title[Heating by H atom impact]{Excitation of Unidentified InfraRed Bands by H atom impact}
\author[R. Papoular]{R. Papoular$^{1}$\thanks{E-mail:
papoular@wanadoo.fr}\\
$^{1}$Service d'Astrophysique and Service de Chimie Moleculaire,\\
CEA Saclay, 91191 Gif-s-Yvette, France}
\begin{document}

\date{Accepted . Received ; in original form }

\pagerange{\pageref{firstpage}--\pageref{lastpage}} \pubyear{2002}

   \maketitle
\label{firstpage}

\begin{abstract}
A model was developed for the excitation of the UIBs by H atom impacts in the Interstellar Medium. It builds upon the fact
 that, in the presence of far UV radiation and hydrocarbon grains, the hydrogen gas will be partially dissociated and the grain 
surface will be partially hydrogenated and partially covered with free carbon bonds. Under such a statistical equilibrium, 
H atoms from the gas will recombine with C atoms at the grain surface at some rate. At each recombination, the H atom deposits
an energy of about 5 eV in the grain. Half of this is directly converted into vibrational excitation, always distributed in 
the same way among the most tightly coupled vibration modes of the grain. Absent frequent grain-grain collisions, the only 
outlet for this energy is IR reemission, part of it in the UIBs, provided the chemical structure of the grains is adequate,
 and the other part in the continuum. The partition only depends upon the grain size, all grains being assumed to have the same 
constitution. Only a fraction, about 0.25, of the grains (among the smallest ones) will contribute significantly to the UIBs.

It is shown quantitatively that H impacts are generally more efficient excitation agents than UV absorption because of the 
overwhelming abundance of hydrogen relative to UV photons. Only very close to young bright stars is this no longer true because photon 
flux then largely exceeds H atom flux. Thus H impacts and FUV absorption are both necessary to understand the variety of observed 
UIB spectra.

The model translates into a small number of equations enabling a quantitative comparison of its predictions with available 
astronomical observations, which have become exquisitely rich and accurate in the last two decades.

\end{abstract}




\section{Introduction}

Infrared radiation from molecules and grains in space is usually or mostly attributed to heating by visible and UV light absorption.
While this may be justified by the ubiquity of light radiation, it must be remembered that the ISM (InterStellar Medium) 
is also permeated with atomic hydrogen. H atoms are also found in abundance in the limbs of molecular clouds illuminated by bright 
young stars (PDRs or photodissociated regions). This paper accordingly
 explores the relative importance of deposition of energy by atomic hydrogen impacts on the same targets. As is well known,
the formation of molecular hydrogen by two isolated H atoms is forbidden  by energy and momentum conservation; it is therefore associated instead
 with recombination in presence of a third body \bf(see Spitzer 1977)\rm. The most readily available such body is a hydrocarbon molecule or grain.
 The latter is made of a 
carbon skeleton to which peripheral H atoms are attached. An incident H radical, if moderately energetic, will attract one of the
H atoms at the grain periphery forcefully enough to capture it and form an H$_{2}$  molecule which readily escapes into space; this is called 
``H abstraction" in surface physics language.
 It will be shown below that this process does not leave much energy in the grain. However, it does leave an unoccupied ``dangling" C-bond.
When the next incident H atom meets that free bond, it is most likely to form a strong chemical bond with the host C atom. This
``recombination" deposits in the grain nearly 4 eV, \bf{half of which is available in the form of kinetic energy to excite the vibrations 
of the hydrocarbon particle (the other half going into potential energy)}\rm. Although this is no more
 energy than is carried by a visible photon, it is totally expendable in vibrational excitation, by contrast with photonic energy, which is 
first delivered to electrons, to relax thereafter into continuum and vibrational radiation. The emission of vibrational bands following 
H-impact excitation is a type of chemiluminescence. \bf The basics of this process were first put forward by Guillois et al. \cite{gui}. 
It may be considered as a special case of the ``radical reactions" previously envisioned
 by Allamandola and Norman \cite{all}; but these authors did not treat any process in particular, dwelling instead on relaxation times and emission 
probabilities. \rm

The present work is also motivated by several astronomical observations which hint at the need to complement the photon flux with another
excitation agent for the emission of UIBs (Unidentified Infrared Bands). One is the fact that the relative intensities of the bands in
 the spectrum of UIBs
do not change notably with the average photon energy of the ambient radiation (Uchida et al. 1998);
besides, neither the paucity of UV radiation (as in 
galaxy M31; Pagani et al. 1999) nor the absence of starbursts (see Haas et al. 2002) precludes UIB emission.

Another hint was the comparison, by  Onaka et al. \cite{ona}, using the IRTS satellite, of UIB emission (5 to 12 $\mu$m) at various galactic
 latitudes throughout the Galaxy: the intensity of emission decreases quickly with increasing latitude, but its spectral profile remains 
unchanged. While there are no hot stars at these latitudes (and therefore much less far UV photons),
 the UIB intensity follows the trend of H atom density, which extends farther 
from the Galactic plane than does molecular hydrogen density (see Imamura and Sofue 1997).

Earlier on, Boulanger and P\'erault \cite{bou88}, studying the solar environment by means of the IRAS satellite, found that, away from 
heating sources
 and molecular clouds, the IR emission from the cirruses of the ISM is well correlated with the column density of HI gas. Pagani et al. \cite{pag}
 reached the
 same conclusion from the study of a large number of sight lines through the near-by Andromeda nebula; by contrast, they found no correlation 
of IR emission with UV flux.

Still another remarkable fact is the occurrence of the UIB intensity peak, in PDRs (photo-dissociation regions) seen edge-on, in between 
the peak of recombination radiation (signaled by H$^{+}$ lines, e.g. Br $\alpha$), on the star-illuminated side (HII region), and the peak of 
H$_{2}$ de-excitation radiation at 2.42 $\mu$m.
This is precisely where maximum atomic H density is to be found. The generic example is the Orion Bar (see Roche et al. 1989), 
Sellgren et al. 1990, Graham et al. 1993). \bf An idealized illustration of the succession of regions and of the distribution of densities and 
radiative fluxes was  given by Guillois et al. \cite{gui}.\rm

\bf Finally, the need for some excitation process other than photon absorption to understand available observations seems to have been felt by 
other workers; e.g. Duley and Williams \cite{dul} reconsidered briefly another type of chemical excitation of IR vibration: the sudden and
violent release, upon mild heating, of potential energy stored in molecules in the form of radicals. \rm

In Sec. 2, I describe the elementary processes which create conditions under which H atom excitation becomes operational. In 
Sec. 3, rate equations coupling the relevant physical quantities are laid down and the statistical equilibrium values of the 
variables are determined. The latter are used in Sec. 4 to determine the energy build up in a grain, under H atom bombardment. Section 5
displays expressions for the UIB reemission power. At this stage, it becomes possible to compare UV and H impact excitation of grains, which is 
done in Sec. 6. Finally, Sec. 7 compares predictions of the chemiluminescence model with observations in various environments and measurements
of different physical quantities. 

\section{Elementary processes}
When an H atom (radical) impinges upon a hydrocarbon target, a number of processes may occur 
(see Lohmar et al. 2009, Cazaux and Tielens 2002, Papoular 2005):

a)Recoil; this is usually nearly elastic, i.e. it involves little energy exchange.

b)Recoil with expulsion of an H atom off the target; this breaking of a C-H bond requires an unusually energetic H projectile ($>$30 000 K).

c)H abstraction, in which the incoming H atom is attracted by a target H atom to form a H$_{2}$ molecule
which readily leaves the target with a high velocity; this leaves a dangling (unoccupied) C bond at the target surface, together with 
a fraction of the C-H bond energy.

d)Surface recombination of the H projectile with a target dangling bond (chemisorption); this may occur over a large range of projectile energies,
 and leaves the whole C-H bond energy in the target.

e) Sticking (physisorption) of the incoming H atom at the target surface, over which it can wander until it goes through (a), (c) or (d).

The steady state coverage of the target depends on all 5 processes but the energy balance is not the same for all, as shown below, \bf{using
 version 7.5 of Hyperchem, the chemical simulation code commercialized by Hypercube, Inc. Earlier versions of this code were used and described in 
detail by, for instance, Papoular \cite{pap00} and \cite{pap05}. Here, I used the semi-empirical PM3 method of molecular dynamics calculation provided by the package.
This method uses quantum-mechanical calculations, supported where necessary by empirical constants from the chemistry laboratory, which are kept up to
 date. In PM3, the empirical part was tailored by comparison with carbon-bearing molecules, so the energies given by this code agree with experiment
 to a few percent, better than necessary for present purposes.}\rm

First, the C-H bond energy was determined in several different structures. This was done by first optimizing the structure of interest (at 0 K), i.e. minimizing
 its potential energy as a function of geometry, and measuring this energy. The H atom was then extracted and the new, higher, potential energy
 measured. The difference is the required bond energy, $E_{b}$.

For H$_{2}$, this is 109.4 kcal.mol$^{-1}$. For an aliphatic chain (-CH$_{2}$-CH$_{2}-$), $E_{b}$=98; in methyl CH$_{3}$, 101 to 106;
 in benzene, 116 to 118, all in kcal.mol$^{-1}$ (1 eV=23 kcal.mol$^{-1}$). \bf{In the following, we take $E_{b}$=4 eV, uniformly, as higher accuracy 
is not warranted in describing the general properties of the present model}\rm.
 Consider first case (d). The target is assumed to be initially at rest at 0 K. In a simplified semi-classical picture, the representative point 
of this state is the bottom of a parabolic potential well \bf as in Fig. 1.\rm When the incoming H atom is captured by a dangling bond, the potential well is suddenly 
depressed by $E_{b}$, so the representative point is now found above the new bottom by the same quantity, with zero KE (kinetic energy). It is
 compelled to slide down the well, rise on the other side, fall down again and repeat this sequence indefinitely. In this new, dynamic, 
state the average KE and PE (potential energy) are both equal to $E_{b}/2$. This energy is initially localized in the new C-H bond in the form of
 large amplitude stretching vibrations, but it quickly spreads to other, closely coupled, vibrations of the structure

 \begin{figure}
\resizebox{\hsize}{!}{\includegraphics{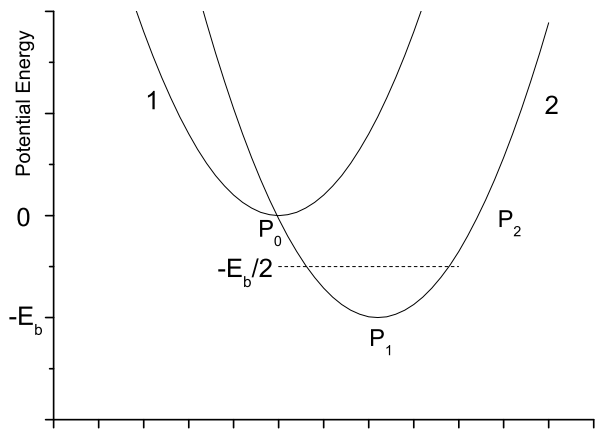}}
\caption[]{Illustrating the kinetic energy gain upon recombination of a gaseous H atom with a C atom of the grain surface, which initially
 presented an unoccupied, or dangling bond (case (d)). The initial state of the molecule is represented by P$_{0}$, on parabola 1. After recombination, 
the representative point oscillates from P$_{0}$ through P$_{1}$ to P$_{2}$ and back, on parabola 2,with average potential energy $-E_{b}/2$ 
and average kinetic energy $E_{b}/2$. In case (b), the system is initially, say, near P$_{1}$. If it is suddenly given enough energy, its representative point
 will rise vertically up to parabola 1, higher than its apex, P$_{0}$, and then oscillate about the latter, having gained again both potential 
and kinetic energy. The horizontal shift of the parabola illustrates the change of stucture accompanying the H capture. } 
\end{figure}

The average life time of this state before energy is lost through IR radiation is usually longer than 1 ms \bf(see Radzig and Smirnov 1980)\rm. 
So the posibility must be 
considered that, before this happens, another H radical is captured by the same target. The representative potential well will again be depressed.
Assume, for simplicity, that the transition to the new, third, state occurs at the moment when the representative point is at its peak height 
in the well with PE=$E_{b}$ and KE=0. Its instantaneous PE above the bottom of the new well will therefore be $2E_{b}$, and the average PE and KE
will rise to $E_{b}$. Thus for each successive H recombination, the target particle gains $E_{b}/2\sim 2$ eV in vibrational energy. If this energy
 were shared equally between, say n=100 atoms of the target, this would correspond to a \bf{vibrational temperature of $2E_{b}/3nk_{B}\sim$170 K.}\rm

Consider now case (c). Here, the gain in PE of the target upon loosing an H atom is partially compensated by the loss of PE associated with the
 formation of a H$_{2}$ molecule, because the H bond energies involved are almost equal. Moreover, the H$_{2}$ molecule carries away vibrational,
 translational and rotational kinetic energies.  Typical values for an aliphatic chain, for instance,
 are, respectively, 19, 15 and 0.3 kcal/mol for the H$_{2}$ molecule, leaving only a few kcal/mol in the target.

For case (b), one can develop the same argument as for (d), except that, here, the target potential well is suddenly raised, instead of depressed,
upon losing an H radical. \bf Assume, in Fig. 1, that the system is given more energy than what is necessary to break a CH bond. After a while, half this extra
 gain in (potential) energy is converted into KE, as in case (d). However, this process clearly requires an unusually energetic H projectile. \rm

 In estimating the overall vibrational energy gain, below, we will therefore neglect
the contributions of recoil and H absraction by comparison with that of H recombination.

\section{Densities in HI regions}

By HI region is meant, here, a region of space permeated with photons energetic enough to dissociate molecular hydrogen without ionizing the atoms. This
 excludes dense molecular clouds but includes, for instance, cirruses in the diffuse ISM. Let $n_{H}, n_{H2}, n_{t}$ be the number densities (cm$^{-3}$) of atomic,
 molecular and nuclear hydrogen, respectively, with $n_{t}=n_{H}+2n_{H2}$. Assume the average velocity of the H atoms is $V$ cm.s$^{-1}$, and 
the flux of dissociating photons is $I$ cm$^{-2}$s$^{-1}$.

 Let $N_{C}$ and $N_{H}$ be the number of C and H atoms in each hydrocarbon grain,
 and $n_{g}$ the number density of such grains (cm$^{-3}$).
Assume, for simplicity, that, in the ISM, the mass ratio between grain and gas is 1/300. Ignoring heteroatoms, which are quite rare in fact, 
 and H atoms which are much lighter, this implies $n_{g}N_{C}m_{C}/n_{t}m_{H}=1/300$, or $n_{g}N_{C}/n_{t}=1/3600$. Since we are interested in H recombination with naked C atoms
(those exhibiting a dangling bond) at the grain surface, we have to distinguish between the naked and the ``H-covered" C atoms. This is
illustrated in Fig. 2, for an aromatic and an aliphatic skeleton.

\begin{figure}
\resizebox{\hsize}{!}{\includegraphics{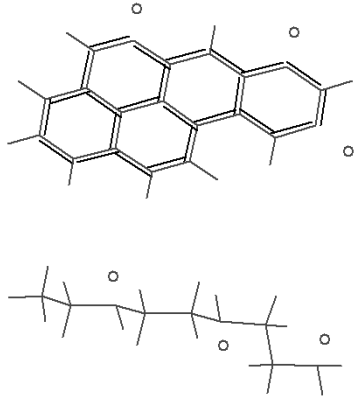}}
\caption[]{Aromatic and aliphatic structures with unoccupied (dangling) C bonds designated by circles.}
\end{figure}

Roughly speaking, a covered C atom may anchor one or two H atoms according to whether the skeleton is aromatic or aliphatic; so there is
 0.5 to 1 covered C atom for every bound H atom. 
Taking 0.75 for simplicity, and
 assuming the carbon skeleton is not too compact, we may therefore write $N_{C}=N+0.75\,N_{H}$ on average. Now, the relative numbers
 of covered and naked C atoms depend on the density of
gaseous H atoms above the grain, and on the various c-s (cross-sections) of the gas-grain reactions. In statistical equilibrium, the number
 of gaseous H atoms captured by a grain is equal to the number of bound H atoms liberated from the grain by H impact. This can be written

\begin{equation}
N(\sigma_{r}+\sigma_{s})=N_{H}(\sigma_{a}+\sigma_{exp}),
\end{equation}

where $\sigma_{r}, \sigma_{a}, \sigma_{exp}$ are c-s's for recombination, abstraction and expulsion, respectively; 
$\sigma_{s}$ is the c-s for recombination after sticking (``indirect recombination"); a factor $n_{H}V$ was
 omitted from each side as it is the same for all reactions, so this equilibrium is independent of the ambient gas density. Then, defining
$y=N_{H}/N$, we get

\begin{equation}
y=\frac{\sigma_{r}+\sigma_{s}}{\sigma_{a}+\sigma_{exp}}
\end{equation}

and, hence, 

$N=\frac{N_{C}}{1+0.75\,y}\,\,,\,\, N_{H}=\frac{yN_{C}}{1+0.75\,y}$.

The c-s's \bf involved in gas-grain interactions \rm have been estimated, computed, modelled or measured (see Cazaux and Tielens 2002, 
\bf Mennella et al. 2002 \rm, Papoular 2005,
 Lohmar et al. 2009). The results concur only approximately. Here, we read
  from Papoular \cite{pap05}, with $V=5\,10^{5}$ cm.s$^{-1}$: $\sigma_{a}=1.9\,10^{-16}$ cm$^{2}$, $\sigma_{r}=2\,10^{-16}$ cm$^{2}$  and
 $\sigma_{s}=\sigma_{exp}=0$. Equation 2 then
gives $y=N_{H}/N=1.05$, indicating that the hydrogen coverage of the target is $N_{H}/N_{C}\sim50\%$. \bf By way of comparison, in a study of gaseous atomic
 hydrogen interactions with carbon grains, Papoular \cite{pap05} found that
the H coverage of hydrocarbon grains varies between $\sim0.2$ and $\sim0.8$, depending mainly on the kinetic temperature of the H gas. In a 
recent study of the hydrogenation of HAC (Hydrogenated Amorphous Carbon) Duley and Williams \cite{dul} estimated the coverage at 0.3-0.5.\rm

Turning to the gas densities, let $\sigma_{d}$ be the H$_{2}$ molecule c-s for dissociation by photons. The rate equations for gaseous atomic and molecular hydrogen densities
can be written

\begin{eqnarray}
\dot{n_{H}}=2n_{H2}I\sigma_{d}-n_{H}N_{H}n_{g}V(\sigma_{a}+\sigma_{s})-n_{H}Nn_{g}V\sigma_{r}+n_{H}N_{H}n_{g}V\sigma_{exp}\,,\\
\dot{n_{H2}}=-n_{H2}I\sigma_{d}+n_{h}N_{H}n_{g}V\sigma_{a}.
\end{eqnarray}

The terms in the r.h.s. of the first of these equations are for photodissociation, abstraction, recombination and expulsion respectively;
  in the second line, the terms are for photodissociation and abstraction respectively. \bf{In statistical equilibrium, the derivatives
 in the l.h.s. must be set to 0.} \rm Defining

$z\equiv\frac{n_{H2}}{n_{H}}\,$,

\bf{and using the second rate equation,}\rm

\begin{equation}
z=\frac{n_{g}N_{H}V\sigma_{a}}{I\sigma_{d}}=\frac{n_{t}V\sigma_{a} y}{3600 I\sigma_{d} (1+0.75 y)}\,,
\end{equation}

\bf{Finally, using $n_{t}=n_{H}+2n_{H2}$}\rm, we obtain

\begin{equation}
n_{H}=\frac{n_{t}}{1+2z}\,,\,n_{H2}=\frac{z n_{t}}{1+2z}.
\end{equation}  

Note that, while $y$ depends only on c-s's, $z$ also depends on the radiative flux and on the velocity of H radicals.
 
As for the dissociating photon flux which enters the rate equations, 
its value \bf{varies considerably with the wavelength\,} \rm in the spectral region of interest (Lyman continuum, 900-1100 \AA{\ }), and so does the dissociation rate
 $\beta=I\sigma_{d}$ (see Draine 1978, Fig. 3, and references therein).
It is therefore usual to replace the latter by an estimated average value, $\beta$, in this wavelength range. We adopt 
$\beta_{0}=5\,10^{-11}$ s$^{-1}$ for
 the Galactic ISM (see Stecher and Williams 1979, Jura 1974). Nearer to hot stars, $I=GI_{0}$, where $I_{0}$ is the Galactic 
FUV (Far Ultra-Violet) flux,
 and $G$ may reach up to $10^{5}$. We shall take $\beta=G\beta_{0}$, assuming that the \bf UV spectral profile \rm does not change as
 the radiation varies in intensity. Using these parameter values, eq. 3 then becomes 

\begin{equation}
z=3.1\,10^{-4}\,\frac{n_{t}}{G}\,,
\end{equation}

Thus, \emph{the ratio of molecules to atoms, z, 
 is proportional to the total number density of H nuclei, and inversely proportional to the UV flux}. 

This simplistic treatment neglects the extinction of dissociating light by molecules and grains along its path, as well as 
self-shielding of the H$_{2}$ molecules against UV due to their excitation out of their ground level (see Draine and Bertoldi 1996, Fig. 8).
 When these effects are included, it is found that, for given density and radiation flux, the light is almost unaffected up to a critical 
column density, beyond
 which it is quickly extinguished while the molecular fraction increases steeply. The elaborate theory helps understanding the earlier observations 
of molecular cloud limbs (see Savage et al.
 1977, Fig. 6), which set the critical column density at a few times $10^{20}$ cm$^{-2}$. The fortunate conclusion for our present
 purposes is that our treatment is acceptable below the critical column density, i.e. in most environments from which
 the UIBs are usually observed to originate: cirruses, molecular limbs and PDRs.

\section{Heating in HI regions}
Assuming the frequency of collisions between grains to be much lower than the frequency of H impacts on a particle, the energy deposited by the latter
accumulates in the particle until it is reemitted in the form of IR radiation, which is its sole outlet.
A fundamental quantity is, therefore, the total kinetic energy deposited per atom during the radiative lifetime. From eq. 2, the total number of
 atoms per particle is

\begin{equation}
N_{t}=N_{C}+N_{H}=w N_{C}\,;\, w=\frac{1+1.75\,y}{1+0.75\,y}.
\end{equation}

From Sec. 2, the vibrational energy deposited by an impinging H atom that recombines on the particle is $E_{b}/2$. The number of such events that
 occur before IR emission is proportional to the radiative lifetime, $\tau_{IR}$, to the ambient H atom influx, to the dangling bond coverage
over the particle surface and to the c-s for direct and indirect recombination. On average, the final (or maximum) vibrational energy \emph{per
 target atom} will be

\begin{equation}
e_{f}=\frac{E_{b}}{2N_{t}}\tau_{IR}Vn_{H}[N(\sigma_{r}+\sigma_{s})+N_{H}\sigma_{exp}]\,,     
\end{equation}
or, \bf using eq. 2 and 3,\rm
\begin{equation}
e_{f}=\frac{E_{b}}{2N_{t}}\tau_{IR}\,V\,N_{C}(\frac{\sigma_{r}+\sigma_{s}+y\sigma_{exp}}{1+0.75y})(\frac{n_{t}}{1+2z}). 
\end{equation}

The characteristic behaviour of this function is determined by the factor ($\frac{n_{t}}{1+2z}$).

As $z$ tends to 0 (high $G/n_{t}$),
\begin{equation}
e_{f}\to\gamma n_{t}\,,\, \mathrm{with} \,\, \gamma=\frac{E_{b}}{2N_{t}}\tau_{IR}VN_{C}(\frac{\sigma_{r}+\sigma_{s}+y\sigma_{exp}}{1+0.75y})\,,
\end{equation}
which is proportional to $n_{t}$ but no longer depends on $G$.

 As $z$ increases beyond 1 (low $G/n_{t}$),
\begin{equation}
e_{f}\to\frac{\gamma n_{t}}{2z}=\delta G\,,\,\mathrm{with}\,\, \delta=900E_{b}\tau_{IR}\beta_{0}(\frac{\sigma_{r}+\sigma_{s}+y\sigma_{exp}}{y\sigma_{a}})
    \,,
\end{equation}
which is proportional to $G$ but no longer depends on $n_{t}$. For the parameters of our numerical example, and $\tau_{IR}=1$ s,
    $\gamma=7.6 \,10^{-11}$ and $\delta=1.2 \,10^{-7}$, if $n_{t}$ is in cm$^{-3}$ and $e_{f}$ in eV.

\section{Emission from HI regions}
For a direct comparison with observations, the total UIB emission per unit volume and time is needed. 
Assume that all the chemical energy
carried by the impinging H atoms is converted into IR emission. But, for the grains to distinctly 
emit the UIBs, they must be \bf neither too small (in which case, they could not 
contain all the functional groups that carry the UIBs), nor too large (in which case most of the energy would go
 into the continuum) \rm : this is demonstrated by emittance mesurements on carbonaceous particles of different
 compositions and sizes (see, for instance, Solomon et al. 1986). 
Thus, only a fraction, $f_{UIB}$, in mass, of the IS carbonaceous grains (as defined by these two constraints) 
must be included in our calculation. Using eq. 8, the UIB emission can then be written

\begin{equation}
P=\frac{f_{UIB}e_{f}N_{t}n_{g}}{\tau_{IR}}=5.2\,10^{-33}\,\frac{f_{UIB}n_{t}^{2}}{1+6.2\,10^{-4}\frac{n_{t}}{G}}.
\end{equation}

Since $e_{f}$ is proportional to $\tau_{IR}$ (see eq. 6), $P$ no longer depends on the latter. Figure 3 displays 
$P/(f_{UIB}n_{t})$, the UIB power emitted per ambient H nucleus.

Again, we distinguish regions of low and high illuminations and write

\begin{equation}
  z\, small: P\sim 2.8\,10^{-4}\,\frac{f_{UIB}w\gamma n_{t}^{2}}{\tau_{IR}}\sim5.2\,10^{-27} f_{UIB}n_{t}^{2}\,,
\end{equation}

\begin{equation}
  z\, large: P\sim 2.8\,10^{-4}\,\frac{f_{UIB}w\delta Gn_{t}}{\tau_{IR}}\sim8.4\,10^{-24} f_{UIB}Gn_{t}\, ,
\end{equation}

where $n_{t}$ is in units of cm$^{-3}$ and $P$ is in W.m$^{-3}$ (choices imposed by observational practice)

In a given IS environment, it may be helpful to determine which of the two cases applies. Equation 8 clearly shows that this depends 
 on whether $z$ is much larger or much smaller than 1/2. The critical condition, $z=0.5$, can be expressed in terms of $G$ and $n_{t}$, using
eq. 5 :

\begin{equation}
G_{c}=6.2\,10^{-4}\, n_{t} ,
\end{equation}

which is displayed in Fig. 4. It is apparent, from eq. 5 and 11, and Fig. 4, that, in the local ISM ($G=1$), dissociation is rather
 high and that eq. 9 and 12 are the better approximations.

\begin{figure}
\resizebox{\hsize}{!}{\includegraphics{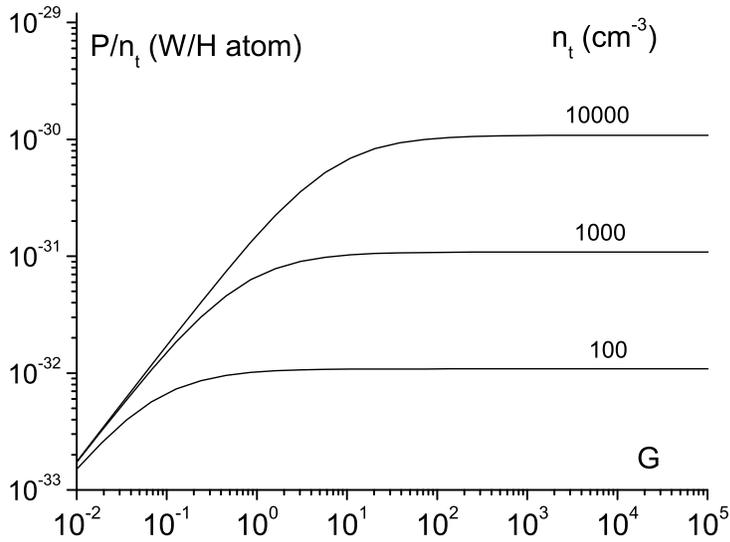}}
\caption[]{The power per H atom reemitted in the IR (eq. 11), for different relevant H nuclei densities, $f_{UIB}=1$ 
and the other parameters and assumptions as stated in the text.
 Note  the saturation at high $G$, and the common asymptote at low $G$, approximated by eq. 12 and 13, respectively.}
\end{figure}

\begin{figure}
\resizebox{\hsize}{!}{\includegraphics{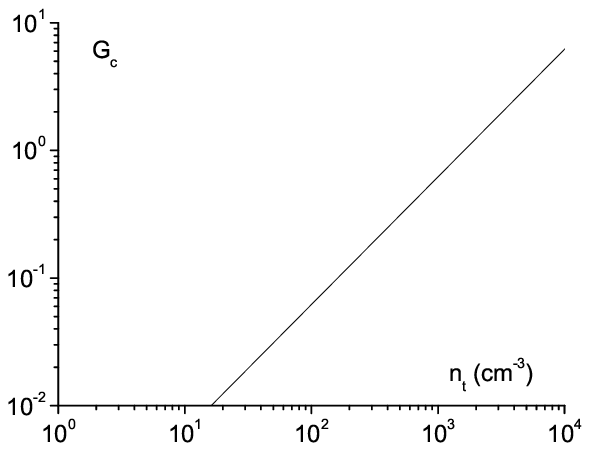}}
\caption[]{The straight line (eq. 14) roughly defines the boundary between two regions: above left, where eq. 9 and 12 apply, and 
 below right, where eq. 10 and 13 apply, approximately.}
\end{figure}

\section{H atoms vs UV photons;\\ UIB vs Total IR emission}
It is instructive to determine the relative contributions of H atoms and dissociative UV photons to grain heating. This is done 
here by considering the ratio of the rates of H atom recombinations (on grains) and UV photon absorptions per unit grain surface, 
\bf multiplied by the corresponding energies deposited in the grain\rm,
\begin{equation}
r=\frac{\frac{E_{b}}{2}N_{H}V\sigma_{r}n_{H}}{h\nu N_{C} I\sigma_{FUV}}\,,
\end{equation}
where $\sigma_{UV}$ is the FUV absorption cross-section per C atom. For C-H bonds, this is $\sim2\,10^{-18}$
 (Turro 2009), while for C=C bonds it rises to $\sim2\,10^{-17}$ cm$^{2}$ (see \,Roche et al. \cite{roc}). For graphite, it is 
$\sim6\,10^{-18}$cm$^{2}$. Let us adopt $10^{-17}$. Also take, as above, $E_{b}/2$=2 eV, $V=5\,10^{5}$ cm.s$^{-1}$, 
$\sigma_{r}=2\,10^{-16}$ cm$^{2}$ and $N_{H}/N_{C}$=0.5. The average energy of dissociating (Lyman continuum) photons will be taken 
to be 12 eV, and $I=7.5\,10^{6}$ photons. cm$^{-2}$.s$^{-1}$ for the diffuse ISM ($G=1$; see Draine 1978 and references therein; 
Mezger et al. 1982). This gives $r=0.11\,n_{H}(cm^{-3})/G$, or, using eq. 4 for $n_{H}$, in the two extreme cases ($z$ large or small):
 
\begin{equation}
G<G_{c}:\,r\sim200\,;
\end{equation}

\begin{equation}
G>G_{c}:\,r\sim\frac{200\,G_{c}}{G}.
\end{equation}

The determining factor in eq. 15 is $Vn_{H}/I$. Thus, \emph{in small FUV fluxes, indirect heating by H atom impacts is much more 
efficient than radiative heating, independent
of $G$; this is essentially because there are so many fast hydrogen atoms that their flux is larger than that of FUV  photons.  
But, in high FUV fluxes, hydrogen is almost totally dissociated, so $r$ decreases because,
 as $G$ increases, $n_{H}$ no longer increases while radiative heating still increases and ultimately becomes dominant.} This conclusion 
holds even if some less energetic photons (1100 to 2400 $\AA{\, }$) are included; it is even strengthened when it is recalled 
that photon absorption is intrinsically less efficient than H impact in producing UIB emission, as some of its energy is diverted 
into ionization and another part into continuum emission.

The ratio (15) can be slightly modified to become the ratio of total UIB emission to total IR emission. To that effect, assume,
 as in Sec. 5, that total conversion
of the chemical energy into UIB emission occurs only in a fraction $f_{UIB}$, in mass, of the carbonaceous grains. Then, the 
numerator of eq. 15 must be multiplied by $f_{UIB}$ in order to represent total UIB emission. On the other hand, note that
  not only dissociating photons (900 to 1100 $\AA{\ }$), but also weaker UV photons (1100
 to 2400 $\AA{\ }$, according to the definition of the Habing UV unit of flux), contribute to the radiative heating 
of grains, and must also be included in the photon flux, $I$; for the latter, we now take $10^{8}\,G$ \bf(based on Fig. 3 of Draine 1978),
 \rm so the denominator now
nearly represents the total IR emission. Expressions 15 to 17 are thus multiplied by $0.075\,f_{UIB}$. This will be compared 
with observations in Sec. 7.1.

\section{Confrontation with observations}
We seek now to compare the predictions of the model with relevant observations.

\subsection{The local diffuse ISM} 
 Consider first the ``diffuse infrared 
emission" from the Galaxy, as measured by Boulanger and P\'erault \cite{bou88} in the solar neighbourhood. 
They compared the IR measurements of the
 IRAS satellite, in its 4 bands, 12, 25, 60 and 100 $\mu$m, with the HI radio emission from the same regions of the
 solar neighbourhood, away from heating sources and outside dense molecular clouds, where $G=1$ by definition.
 Noting that the radio emission is proportional to the H atoms number density, they 
deduced the band emission per H atom, in each band, and found 1.1, 0.7, 1.1 and 3.2, respectively, in units of $10^{-31}$
 W per H atom, for a total of 6.1. Boulanger and P\'erault \cite{bou88} also derived an estimate of the emission
 per H atom of the diffuse ISM,
 in the UIB range (2 to 15 $\mu$m): $\sim1.5\,10^{-31}$ W.H atom$^{-1}$. From these data, $f_{UIB}$ can be deduced. For 
the contribution of UV 
photons to direct grain excitation is negligible in the diffuse ISM (see Sec. 6), so both UIB and total IR emissions are
excited by H impact (in the present model). As a consequence, their ratio is precisely equal to $f_{UIB}$,
according to our definition of the latter (Sec. 5), so

\begin{equation}
f_{UIB}=1.5/6.1\sim0.25\,.
\end{equation}

 This considerable fraction is a necessary consequence of the assumption that UIB and other IR emissions are excited by 
the same agent. This constraint can be met with H atom excitation because even relatively large grains can be excited;
 the upper limit to their size is several thousands of C atoms per grain , before the continuum confiscates most of the 
chemical energy. In fact, the grain efficiency in exciting UIBs
 must be a decreasing function of grain size, so the transition between UIB-emitting and non-UIB-emitting must be gradual rather
 than a sharp cut-off.

 Pagani et al. \cite{pag} made a similar study of the diffuse ISM of the Andromeda Nebula, M 31. They considered the signal
 from  the LW2 (5-8 $\mu$m) filter of ISOCAM 
on board the ISO satellite as a tracer of UIBs and found that \emph{it is extremely well correlated, 
across the galaxy, with the distribution
 of neutral gas as mapped by the HI (21 cm radio) and CO(1-0) signals (\rm{$n_{t}$ in our model}), which they also plotted. 
By contrast, the correlation is poor with the ionized
 gas as seen through its H$\alpha$ emission, and non-existent with the UV emission}. They conclude that the mid-IR 
emission can be excited by the ambient visible and near-IR radiation. Here, I argue that, if that were the case, 
there would still be a much tighter correlation with the young stars of the nebular ring than shown by the maps. 
Instead, I set out to interpret quantitatively the findings of Pagani et al. \cite{pag} by means of the present model.

Pagani et al. analyzed their images, pixel by pixel, and plotted their measurements as a graph 
of I(LW2) as a function of the column density, N(H), in their Fig. 6. From this, they deduced the regression line
\begin{equation}
I(LW2)= 2.24\,10^{-22}\,n_{t}L-0.23\,,
\end{equation}

\bf where $n_{t}L=N(H)$, \rm and I(LW2) is in mJy/pixel and $n_{t}L$, in cm$^{-2}$. Given the size of the pixel (6"x6"), and neglecting the
 ordinate at the origin, -0.23, we deduce
\begin{equation}
<\rho>\equiv\frac{I(LW2)}{n_{t}L}=3.5\,10^{-32}\,\rm {W per Hatom}.
\end{equation}

Since the ISOCAM images cover mainly the general (diffuse) ISM of M 31, this number can be compared with the
 value, 1.5$\,10^{-31}$, measured by Boulanger and P\'erault \cite{bou88} in our Galaxy, in the UIB range (2 
to 15 $\mu$m). They are of the same order of magnitude, but the signal from M 31 is weaker by a factor $\sim$4. This must be due,
 in large part, to the fact that
the LW2 filter covered the range 5-8 $\mu$m, \bf thus excluding the strong 11.2 and 12.7 $\mu$m bands, \rm while the measurement of the Galaxy covered the larger range 2-15 $\mu$m. 
 The remaining difference is discussed next.

Figure 6 of Pagani et al. exhibits a significant dispersion about the regression line. This is quantified in
 their Fig. 9, which plots $\frac{I(LW2)}{n_{t}L}$ (\bf our $<\rho>$ \rm) as a function of the FUV flux in a band of 15 nm width around 200 nm. When
translated in the units used here, their measurements extend over \bf the ranges 
0.25 to 0.8$\,10^{-17}$ erg.cm$^{-2}$.$\AA{\ }^{-1}$.arcsec$^{-2}$ for the FUV flux, and
 0.32 to 4.75$\,10^{-32}$ W/ H atom, for $<\rho>$. \bf Boulanger and P\'erault \cite{bou88}, had also found considerable patchiness in the 
IR emission from the solar neighbourhood in all directions (their Fig. 3). This was associated with the notion of ``cirrus" imagined
 by Low et al. \cite{low} by analogy with this type of tenuous clouds in our atmosphere.

In the present excitation model, this dispersion can be interpreted in terms of H density differences between regions having
 the same UV illumination. In order to \bf justify \rm this proposition, we return to  eq. 11, 
which gives the exact expression of the power reemitted in the UIBs per 
unit volume of space, insert 0.25 for $f_{UIB}$ and divide by 2 to take into account the smaller band width of the LW2 filter.
 The result can be cast as

\begin{equation}
\rho=6.5\, 10^{-34}\frac{n_{t}}{(1+2z)}\,\, \rm{W per Hatom}\,
\end{equation}
where $z=3.1\,10^{-4}n_{t}/G$, \bf as in eq. 5, \rm and $n_{t}$ is in cm$^{-3}$.

Clearly, $\rho$ varies with $n_{t}$ and $G$. As an example, take the minimum measured UV flux in Fig. 9 \bf of Pagani et al. \cite{pag} \rm
(for which the 
dispersion of plotted points is greatest): 2.4 $10^{-18}$ erg.s$^{-1}$.cm$^{-2}\,.\AA{\ }^{-1}$.arcsec$^{-2}$, which is close to the
 value in the solar neighbourhood: 2.1 10$^{-18}$ (see Draine 1978). For lack of better information, assume
that the ratio of dissociating UV flux to the flux at 200 nm is the same in M 31 as in the solar neighbourhood. Then, for the selected points, 
$G\sim1$, according to our definition of $G$ in Sec. 3. For the corresponding minimum and maximum values of $\rho$, eq. 19
gives, respectively, about 5 and 77 H at.cm$^{-3}$. For the average $\rho$ (eq. 20), $n_{t}=56$ cm$^{-3}$. These calculations 
implicitely assume that every measured IR power emanates from a limited region of space (a cirrus), with a relatively homogeneous density 
distinctly different from that of the environment (see also Sec. 7.5).  

Boulanger and P\'erault \cite{bou88} also derived an estimate of the emission per H atom of the diffuse ISM, within 1 kpc of the Sun, 
in the UIB range (2 to 15 $\mu$m): $\sim1.5\,10^{-31}$ W.H atom$^{-1}$. In this case, eq. 21 gives $n_{t}=120$ cm$^{-3}$. For densities in 
this range and $G=1$, $\rho$ is much more sensitive to variations of $n_{t}$ rather than in $G$. So, although the FUV content 
in M 31 may be lower than in the solar environment, as suggested by Pagani et al. \cite{pag}, the weakness of the measured $\rho$
 may, in the present model, be due to the relative paucity of hydrogen rather than to that of the FUV flux.

\subsection{PDRs and Cold molecular clouds}
When a young, bright star (e.g. class O or B), happens to be in the vicinity or inside a denser (and hence cooler) cloud, its FUV
radiation digs into the cloud to form an HII region (signalled by ionic radiation or H recombination radiation; e.g. Br $\alpha$,
 P $\alpha$), followed farther
 by an HI region (PDR proper; essentially H radicals, signalled by centimeter-band radio emission), then by a thin layer of radicals recombining
 into excited molecules (signalled by several near IR
 lines; e.g. 2.42 $\mu$m) and, farther still, by the new edge of the molecular cloud (signalled by CO molecular lines). This sequence
has often been described; for a recent analysis, see Habart et al. \cite{hab}. For our present argument, it is essential to note
 that, in this type of environment, the UIB are observed to come from the intermediate, HI region and the adjacent H$_{2}^{*}$ region 
\bf(where vibrationally excited H is observed)\rm: 
see, for instance, Roche et al. \cite{roc} for the 11.3- and 12.7-$\mu$m bands, Sellgren et al. \cite{sel90} for the 3.3-$\mu$m band, all 
observed in the Orion Bar; An and Sellgren \cite{an}, studying NGC 7023, compared the maps of the 3.3-$\mu$m band with its adjacent, 
underlying, IR continuum
as well as the 1-0 S(1) line of excited H$_{2}$; Giard et al. \cite{gia} studied M17 at 3.3 and 10 $\mu$m, at 6 cm and in the C$^{18}$O 2-1 band.
This is a very strong indication of a link between H atoms and UIB excitation, especially when it is remembered that, in the present model,
those molecules that are formed after an H atom impact on a grain, leave the grain surface with a large amount of vibrational excitation 
(see Sec. 2) and that the H$_{2}^{*}$ near IR emission is therefore generated  at the interface between the HI region and the edge of the molecular cloud.

Boulanger et al. \cite{bou98} later analyzed the data collected by the ISO satellite from several such locations.
They were thus able to quantitavely determine the brightnesses of the main UIBs, at 6.2, 7.7, 8.6, 11.3 and 12.7 $\mu$m, 
and this for several sources for which they also determined the illumination $G$. They found a common trend for  all five 
bands: an initial rise in intensity as a function of illumination, followed by a progressive levelling-off for 
$G>1000$ (their Fig. 3). The brightness of the 7.7 $\mu$m band, for instance,  starts at $10^{-7}$ W. m$^{-2}$.sr$^{-1}$  for $G=1$ 
(the Chamaeleon cirrus cloud in the dffuse ISM), and terminates at 6.4 W. m$^{-2}$.sr$^{-1}$  for $G=10^{5}$ (the Omega nebula, M 17). 

The single UV photon excitation (also referred to as transient
 or stochastic heating \bf in the PAH hypothesis\rm ) expects a continuous rise of band brightness with $G$, and the levelling-off is tentatively explained by
grain destruction. The present model predicts an initial linear rise in incompletely dissociated regions followed by a  levelling-off
as a result of high dissociation, and limited amount, of available H gas.

For a more quantitative comparison of our model with observations, consider
 the emission $P$ given here by eq. 11. This IR power is emitted in 4$\pi$ sr. The IR surface brightness of a 
source of depth $L$ along the line of sight is,
 then, $I=LP/2\pi$. The results of astronomical measurements
are usually expressed in the form $W(\lambda)=\lambda I_{\lambda}$, where $I_{\lambda}=\frac{dI}{d\lambda}$. Now, the 
spectrum of a finite material structure (or small grain), as 
opposed to that of a homogeneous medium of infinite extent, is expressed in terms of its discrete line intensities, $A_{\lambda}$. 
There are $3N_{t}-6$
of these for a total of $N_{t}$ atoms (one for each vibrational mode); they can be measured (absorptivity) or 
calculated (electronic polarizability). If the line distribution is sufficiently dense, an adjacent averaging
over $\Delta\lambda$ makes sense, giving a continuous distribution

 $a(\lambda)=\frac{\Sigma_{\Delta \lambda}A_{\lambda}}{\Delta\lambda}.$

Assuming the excitation energy to be uniformly distributed among all vibrational modes , one can write

$\frac{I_{\lambda}}{I}=\frac{a(\lambda)}{\Sigma_{\lambda}A_{\lambda}}$ ,

and the observed astronomical quantity becomes

\begin{equation}
W(\lambda)=\frac{\lambda a(\lambda)LP}{2\pi\Sigma_{\lambda}A_{\lambda}},
\end{equation}

where $L$ is in meters, $P$ in W.m$^{-3}$ and $W$ in W.m$^{-2}$.sr$^{-1}$. Finally, a typical value of the pure number
\begin{equation}
\phi(\lambda)\equiv\frac{\lambda a(\lambda)}{\Sigma_{\lambda}A_{\lambda}},
\end{equation}

can be derived from a bright source, like NGC1482 (Smith et al. 2007), for each of the main bands. This was done
 here by interpolating the flux spectrum with 600 points between 5 and 20 $\mu$m, so that $\Delta \lambda=2.5\,10^{-2}\,\mu$m.
 For the 7.65- and 11.3-$\mu$m bands, for instance, one thus obtains 1.4 and 2.1 respectively. 

Inserting $\phi(\lambda)=1.4$ in eq. (22), together with expression (12) for $P$, and $f_{UIB}=0.25$ (as in Sec. 7.2), we obtain

\begin{equation}
W(7.65)=2.9\, 10^{-30}\frac{Ln_{t}^{2}}{1+6.2\,10^{-4}\frac{n_{t}}{G}}\,.
\end{equation}

Thus, for $W(7.65\,\mu$m) to comply with the observed value of 10$^{-7}$ with $G=1$, $Ln_{t}^{2}$ should be
 3.3$\,10^{22}$ cm$^{-5}$ with, for instance,
$L$=1 pc and $n_{t}=100$ cm$^{-3}$, which seems reasonable. For this choice, $z\sim3\,10^{-2}$ and, in Fig. 4, the representative
 point falls above the critical line, which would
 justify the use of approximations 9 and 12.

It is tempting to use the relation between hydrogen column density and visible extinction along the same line of sight (if available):

\begin{equation}
Ln_{t}=2\,10^{21}\,A_{V} (cm^{-2})
\end{equation}

 (see Bohlin et al. 1978). Here, $Ln_{t}$=3$\,10^{20}\,$, so $A_{V}=0.15$; but, in the case of Chamaeleon, Laureijs et al. 
\cite{lau} measured $A_{V}$=1. The discrepancy may be due to a number of causes, the most probable of which is that, 
for this value of A$_{V}$, the molecular hydrogen fraction in the column density
may be quite high due to light extinction by molecules (see Savage et al. 1977, Fig. 6; Draine and Bertoldi \cite{dra96}, Fig. 8);
 in this case, our expression 3 would underestimate $z$, thus leading to a lower estimate of $n_{t}$.

At the high $G$ end of Boulanger et al.'s observations, for M 17 with $G=10^{5}$, $W$(7.7$\mu$m)=6$\,10^{-4}$ W.m$^{-2}$.sr$^{-1}$.
Equation (26) then gives $Ln_{t}^{2}\sim2\,10^{26}$ cm.cm$^{-6}$. Here, the geometry of the emitting region must be very different than that of a thin
 cloud, because the very strong illumination by a young, bright, star near by drives a shock in a thick cloud, thus considerably increasing the 
density in a region of limited extent: a ``bar" whose thickness along the direction to the star is a small fraction of a parsec
 (see, for instance, the extensive study of the Orion Bar by Bregman et al. \cite{bre}). A more likely order of magnitude for $L$, therefore,
 is now 0.1 pc, for which one gets $n_{t}\sim2.7\,10^{4}\,$cm$^{-3}$, again within reasonable range for this type of object. 

More recently, analyzing the data of the \emph{Spitzer} satellite, Smith et al. \cite{smi07} produced a large number of good quality UIB
 spectra from star-forming galaxies. The peak brightnesses of the main bands cover the range between 10$^{-6}$ and $10^{-4}$ 
W.m$^{-2}$.sr$^{-1}$, similar to the Galactic data of Boulanger et al., confirming the remarkable uniformity of the UIB \bf spectral profiles, 
independant of their intensity. \rm

\subsection{The ratios of UIB intensities}
A crucial result of Boulanger et al.'s \cite{bou98} observations, among others, is that the relative intensities of the main
 UIBs do not vary much, nor systematically, over a $G$ range
of $10^{5}$. This insensitiveness was noticed by several authors, observing various environments (out of 
Galactic plane, other galaxies, etc.): see, for instance, Murakami et al. \cite{mur}, Chan et al. \cite{cha}, Mattila et al. \cite{mat99}, 
Kahanp\"a\"a et al. \cite{kah}, Smith et al. \cite{smi07}.
It is in agreement with H-impact excitation, which, in nanoparticles, is followed by non-thermal, quasi uniform, redistribution
 of the same chemical energy (5 eV) among those vibrational modes of the target that are strongly coupled. By contrast, the insensitiveness 
to photon flux is at variance with what one would expect from single UV photon excitation as noted by Sellgren et al. \cite{sel90}.
 In this model, the grain temperature is supposed to decrease
with the photon energy so the relative intensities of shorter-wavelength bands would also decrease; the UIBs should even fade out
precipitously for illuminating star temperatures below $\sim10\,000$ K, as the FUV content of the illumination is then
too small. But this was not borne
out by their observations of several Reflection Nebulae illuminated by stars with temperatures ranging from 5000 to
 21000 K (see also Uchida et al. 1998).

Mattila et al. \cite{mat99} also remarked that their and others' observations implied that the UIB carriers are very resistant to 
different environments (RNe, PNe and HII regions). 

None of the above is incompatible with the small spectral variations in position, intensity or width, which led to the
 classification of UIB spactra in different
classes (see Peeters et al. 2002, Tokunaga 1997). In some cases, the variations are more radical, such as for the 11-13 $\mu$m 
massif across NGC7023 (Boulanger et al. 1999). Still more striking is the detection or non-detection of the 
3.3-$\mu$m band, independent of illuminating star temperature or ambiant UV content over 4 orders of magnitude of $G$ 
(Smith et al. 2004). This poses a common problem to all models, but appears to be linked
with particular chemical processes or environments, rather than with the excitation process; as such, it falls outside the scope
 of this work.

\subsection{Band-to-continuum contrast ratio}
In most UIB spectra, an underlying continuum is clearly apparent, rising towards long wavelengths and often very weak at short 
wavelengths (see, for instance, Cesarsky et al. 2000, Peeters et al. 2004, Zhang and Kwok 2010).
From their analysis of measurements with the ISO satellite in the 6-12 $\mu$m band and at 100 $\mu$m, Kahanp\"a\"a et al. \cite{kah} 
concluded that both emissions were strongly correlated at small and large scales, which implies a strong connection (origin,
 chemical composition, environment) between the small grains responsible for the former, and the large ones, responsible for the
 latter. Mattila et al. \cite{mat99} arrived at the same conclusion from their study of the disk of NGC 891, which they observed
 at 8 wavelengths characteristic of different ISM components. This conclusion is in line with the present treatment, as well as 
with the roughly constant ratio of UIB to total
 IR emission (Sec. 7.1). However, it seems to be at odds with the large variations of the spectral band-to-continuum ratio, or 
contrast, $R$,
from object to object (see also Boersma et al. 2010). If the grain composition does not change with size, this may be explained as follows.

The continuum brightness contribution to the band at $\lambda$ can be written

\begin{equation}
\beta=\epsilon(\lambda)\lambda\,\frac{B(\lambda,T_{g})}{2\pi}L\,n_{g}S_{g}\,
\end{equation}

where $\epsilon$ is the grain emissivity, $B$ is Planck's black-body law (W.m$^{-2}$.$\mu$m$^{-1}$), $L$ the object depth along
 the line of sight (m), $n_{g}$ the grain number density (m$^{-3}$), and $S_{g}$ the grain optical c-s (m$^{2}$). 

The grain temperature is estimated by assuming thermal equilibrium and equating the power absorbed by a grain from the 
illuminating flux $F$, to the power reemitted by the same grain in the IR. For simplicity,
 assume the grain absorptivity 
(emissivity) varies as $\lambda^{-1}$ and that all the reemitted power 
is at the same wavelength, viz. the peak of the BB (Black Body)
 emission, $\lambda_{g}=3000/T_{g}$, and is given by the Stefan-Boltzmann law,
\begin{equation}
\epsilon(\lambda)\sigma_{SB}\,T_{g}^4\,n_{g}S_{g}\,,
\end{equation}
where $\sigma_{SB}=5.7\,10^{-12}\,$W.cm$^{-2}$.K$^{-4}$.

 Also assume the flux that heats the large grains is $F=GF_{0}$, where $G$ is the same as
above, but $F_{0}$ is much larger than the dissociating flux, in order to include lower energy photons. Following Boulanger and 
P\'erault \cite{bou88}, we take $F_{0}=10^{-9}\,$W.cm$^{-2}$. Also, defining the photon wavelength as 0.1 $\mu$m, one gets

$\alpha_{IR}=\alpha_{UV}T_{g}/3\,10^{4}$.

Then, equating absorbed and reemitted powers, the grain diameter
 and its emissivity (at the reference wavelength) factor away, and it is easily deduced that

\begin{equation}
T_{g}\sim17\,G^{0.2}\,,
\end{equation}

Equation (26) still requires an estimate of the emissivity, $\epsilon$; it can be obtained from the consideration of the total far IR emission
 measured by Boulanger and
 P\'erault \cite{bou88} in the IRAS 60- and 100-$\mu$m bands: $4\,10^{-31}$ W per H atom. This is to be equated with the 
Stefan-Boltzmann law. Now, Spitzer \cite{spi} estimated the grain c-s at $\sim10^{-21}\,$cm$^{2}$
 per H atom. Assuming the fraction associated with carbonaceous dust to be $f_{c}=0.5$, then
 $n_{g}S_{g}=0.5\,10^{-21}\,n_{t}\,$cm$^{2}\,$. Finally, taking $T_{g}=20\,$K, we obtain $\epsilon=10^{-3}$ at
 the peak of the black body emission, 150 $\mu$m and, for an opacity in $\lambda^{-1}$,

\begin{equation}
\epsilon(\lambda)=10^{-3}(\frac{150}{\lambda})\,.
\end{equation}

Using the Spitzer result again to express $n_{g}S_{g}$ in eq. 26, the latter becomes
\begin{equation}
\beta=\epsilon(\lambda)\lambda\,\frac{B(\lambda,T_{g})}{2\pi}\frac{L\,n_{t}}{2\,10^{19}}\,
\end{equation}

Finally, the contrast becomes

\begin{equation}
R=\frac{W}{\beta}=2\,10^{19}\frac{P\phi}{\epsilon(\lambda)\lambda\,B(\lambda,T_{g})n_{t}}\,
\end{equation}
where $P$ is taken from eq. 11 and $\phi$, from eq. 23. As an illustration, Fig. 5 plots $T_{g}$ as a function of $G$, as well
as $R$ for $\lambda=7.65\,\mu$m and 11.3 $\mu$m, both for $n_{t}=10^{3}$ cm$^{-3}$. Our model predicts that, for most
 environments in the ISM,
$G$ and $n_{t}$ are such as to deliver high contrasts, as usually observed. It also predicts that the contrast decreases 
as the UIB wavelength increases, as observed, except for the diffuse ISM (see Onaka et al. 1996): this is because, 
there, hydrogen is dissociated although the illumination is so low that the larger grains are very cold and the continuum very weak.
 Finally, the model predicts that the transition from
high to low contrast is very swift as $G$ increases, due to the steep short wavelength edge of Planck's law. As a result,
the UIBs paradoxically ``disappear" at high illuminations! In the present model, this is not due to their destruction,
 but to the fact that the underlying continuum quickly increases, even as the UIB emission levels off because all available
hydrogen is already dissociated (cf. Fig. 3). 

\begin{figure}
\resizebox{\hsize}{!}{\includegraphics{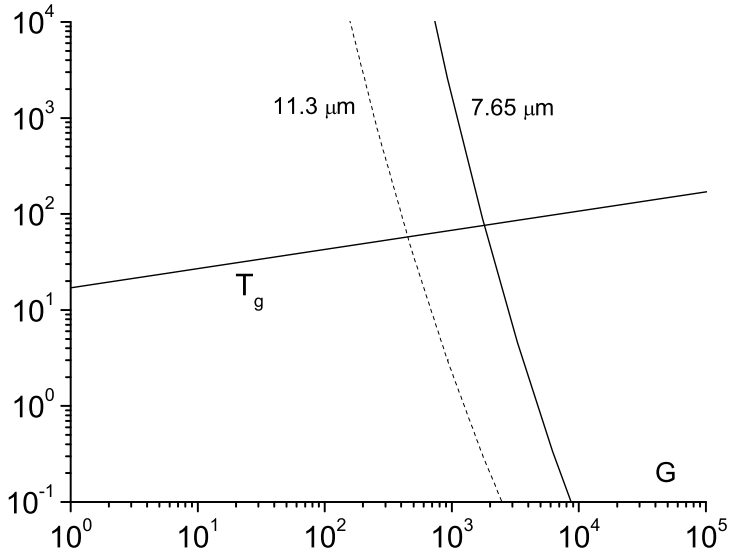}}
\caption[]{As a function of illumination, $G$: the grain temperature, $T_{g}$; the band-to-continuum ratio $R$, at 7.65 and 
11.3 $\mu$m (full and dashed lines, respectively) all three for $n_{t}=10^{3}$ cm$^{-3}$.}
\end{figure}

Observational examples of this behaviour of the band-to-continuum ratio can be found in Giard et al. \cite{gia} 
(3.3-$\mu$m UIB, 10-$\mu$m 
continuum) and Cesarsky et al. \cite{ces96} (6.2-$\mu$m UIB, 16-$\mu$m continuum), both observing M 17.

The predictions of our model concur with the conclusions of An and Sellgren \cite{an}, based on their observations of the
 Reflection Nebula NGC 7023
 with high spatial resolution, in the 3.29-$\mu$m UIB and its neighboring continuum at 2.18 $\mu$m. They find the two
 emissions to coexist in a vast HI region between the HII region enclosing the illuminating star, and the thin H$_{2}^{*}$
filament at the edge of the molecular cloud. However, the intensity distributions are distinctly different, with the 
3.29-$\mu$m peaking within the filament, while the 2.18-$\mu$m peaks at about half this distance from the star. Moreover,
 they measured a feature-to-continuum ratio peaking at
 $\sim23$, in the filament, and decreasing with the distance, $r$, from the illuminating star as $r^{2.1}$, i.e. 
roughly as $G^{-1}$ (here, $G$ is proportional to the star flux heating the grains). Remarkably, 
this trend is very smooth, in stark contrast with the very dissimilar spatial distributions of the two emissions. 
As noted by the authors, both these 
facts are hardly understandable if both emissions are prompted by absorption of the star radiation. On the other hand,
 the present model predicts just this behaviour of the band contrast. 

However, Fig. 5 suggests a much faster variation with $G$. Note that the $r^{2.1}$ trend is only a regression line
over a large set of highly dispersed points, covering only a small range of $G$: a factor 2, which suggests high density heterogeneity.
So the discrepancy may be due to the lines in Fig. 5 being drawn for constant $n_{t}$. But it could also be due to a number of assumptions
 made for the sake of simplicity:

 -assuming too tight relationships between the densities of large grains
 (emitting the continuum), UIB-emitting grains and H radicals (Sec. 3). These relationships hold best in the diffuse ISM, and 
may be invalidated in the presence of very intense UV radiation able to drive a shock front into the H cloud.

-assuming that the illuminating spectrum does not change with $G$, while, in fact it varies considerably from the diffuse ISM 
to the neighbourhood of a bright star.

 \subsection{The Infra-Red ``Excess"}

Ever since Pipher \cite{pip} discovered the Diffuse IR Emission (DIRE) originating in the Galactic plane, the problem of 
 its excitation has been open for discussion. Based on observations by Low et al. \cite{low} and Rouan et al. \cite{rou},
respectively, Mezger \cite{mez78} estimated the total DIRE at $L_{IR}\sim 2.4\,10^{36}\,$W, while Ryter and Puget \cite{ryt}
 set it at 7.5$\,10^{36}\,$W. Given reasonable star formation rates, it has been hard to argue that this emission can be 
accounted for by the luminous O stars alone. Mezger estimated that O stars fuelling radio HII regions contribute 
only 20 \% of the total. He also concluded that, for UV absorption and IR reemission by the Galactic grains in the diffuse medium to 
account for $L_{IR}$, these grains should be able to absorb 8 times more UV photons than they can do in fact (this is the IR ``excess").

Section 6 above suggests that H atom impacts might do the job. In order to explore this possibility in the present 
model, note that the total number of H atoms in the Galaxy was
 estimated by Ryter and Puget \cite{ryt} and Mezger \cite{mez78}, again excluding the Galactic Center, at $5\,10^{9}\,$M$_{\odot}$,
 which amounts to $5.3\,10^{66}\,$ H atoms. Using the values of $L_{IR}$ mentioned above, we deduce the total IR power reemitted 
per H atom, $\rho$=4.5 and 14 $10^{-31}$ W/H atom. These are comparable to the value deduced from the IRAS data by Boulanger and 
P\'erault \cite{bou88}: 6.1 $10^{-31}$. Now, $L_{IR}$ must be compared with the value 
predicted by our model, i.e. eq. 11, which gives the total IR power reemitted per unit volume of space, if $f_{UIB}$ is replaced by 1.
 Here, as in Sec. 7.1,
we deal with a tenuous environment, so approximation (12) may be used instead. Integrating this quantity over the volume of the
 Galaxy (excluding star forming regions), we obtain 

\begin{equation}
\int 5.2\, 10^{-33}\,n_{t}^{2}\,\rm {dV}\,,
\end{equation}

where $n_{t}$ is in cm$^{-3}$. The total volume, V, of the Galaxy is the ratio of the total number of H nuclei divided by the average
density of the latter. This is about 1 cm$^{-3}$ (see Spitzer 1977), so V$\sim5\,10^{66}$ cm$^{3}$. The root mean square density can then be deduced:

\begin{equation}
<n_{t}^{2}>=\frac{L(IR)}{ 5.2\, 10^{-33}}\,\rm {V}\sim200\,\rm{cm}^{-6}\,,
\end{equation}

where $L(IR)$ was set at $5\,10^{36}$ W. The square root of this is 14 cm$^{-3}$, which is
typical of clouds containing optically thin H$_{2}$ in the DISM (see Jura 1975); the model thus proves to be sound.
The difference between $<n_{t}^{2}>$ and $<n_{t}>^{2}$ is a measure of the patchiness of the diffuse ISM.

\section{Summary and Conclusions}
A model was developed for the excitation of the UIBs by H atom impacts in the Interstellar Medium. It builds upon the fact
 that, in the presence of far UV radiation and hydrocarbon grains, the hydrogen gas will be partially dissociated and the grain 
surface will be partially hydrogenated and partially covered with free carbon bonds. Under such a statistical equilibrium, 
H atoms from the gas will recombine with C atoms at the grain surface at some rate. At each recombination, the H atom deposits
an energy of about 5 eV in the grain. Half of this is directly converted into vibrational excitation, always distributed in 
the same way among the most tightly coupled vibration modes of the grain. Absent frequent grain-grain collisions, the only 
outlet for this energy is IR reemission, part of it in the UIBs, provided the chemical structure of the grains is adequate,
 and the other part in the continuum. The partition only depends upon the grain size, all grains being assumed to have the same 
constitution. Only a fraction, $f_{UIB}\sim0.25$, of the grains (the smallest ones) will contribute significantly to the UIBs.

It is shown quantitatively that H impacts are generally more efficient excitation agents than UV absorption because of the 
overwhelming abundance of hydrogen relative to UV photons. Only very near young bright stars is this no longer true because photon 
flux then largely exceeds H atom flux. Thus H impacts and FUV absorption are both necessary to understand the variety of observed 
UIB \bf spectral intensities.\rm

The model translates into a small number of equations enabling a quantitative comparison of its predictions with available 
astronomical observations, which have become exquisitely rich and accurate in the last two decades. Section 7 collects some of 
the most crucial tests of the model:

- It predicts UIB emission even far from star bursts and in poor-FUV environments.

- Since the excitation agent and the deposited energy are always the same, UIB reemission is independent of the temperature 
of the neighbouring illuminating stars.

- In PDRs, it becomes possible to understand the observed very near spatial coincidence of HI and H$_{2}^{*}$ emission with 
the peak of the UIBs.

- The model predicts the measured average power in the UIBs (2-15 $\mu$m) emitted per ambient H atom (in statistical equilibrium
 in the solar neighbourhood),
 1.5 $10^{-31}$ W, provided that 1/4 of the cosmic carbon be locked in UIB-emitting grains, and the average H nuclei density  
 be about 100 cm$^{-3}$ in the average emitting cloud.

- Agreement between computed and measured UIB intensities is obtained for a given FUV illumination, provided the unknown H column 
densities are given adequate values, which are found to be quite plausible. This is true even for M 31, which is presumed to be 
poor in FUV flux.

- The assumption of a common carrier for the UIBs and the underlying continuum makes it possible to quantitatively account for the 
observed large variations of the band-to-continuum contrast ratio: they are simply due to the 
steep variation of Wien's law, at a given wavelength, for small variations of illumination.

- The ``IR excess" paradox, posed long ago by the comparison of the overall Galactic IR emission with the UV flux available for grain
 excitation, is resolved quantitatively by H atom excitation, provided the r.m.s. density of H nuclei throughout the Galaxy be 
of order 200 cm$^{-6 }$.

\bf In order to allow quantitative comparisons to be made with observations, a number of assumptions were necessary. The most restrictive
 one is the neglect of UV extinction and self-shielding of hydrogen; the calculations are progressively invalidated as column densities 
along the sight lines exceed a few times 10$^{20}$ cm$^{-2}$. Fortunately, this is not the case for most of the mesured UIB spectra. 

 For lack of relevant experimental results or explicit values from theoretical computations, I relied heavily on numerical chemical 
simulation to obtain the values of cross-sections of grain-gas interactions. As the heating is proportional to the C-H recombination c-s (eq. 8),
 the latter needs measurement or at least confirmation by other means. It was also assumed
 that the importance of the dissociating UV flux relative to total UV does not depend on the total UV flux; this only impacts the degree of
 dissociation of hydrogen, and most of our calculations are weakly sensitive to it. None of these assumptions notably affects the general 
nature and performance of this model.

The notion of a fraction $f_{UIB}$ should not be taken to correspond to definite lower and upper size 
limits; for the transition between emitting and non-emitting grains is progressive and likely to depend on grain structure and 
composition. Some of the physical processes involved here are evidenced by molecular dynamic simulations. Even for a 100-atom molecule, an 
important broadening mechanism is already at work and directly observable in chemical simulations:
 this is the interaction between modes and constant redistribution, between them, of the initially deposited energy. This widens lines into
 bands and confers wings to 
each line, so ``plateaus" appear below ``bands" in crowded spectral regions, and even an underlying continuum is visible (see 
Papoular 2000). This effect is observed to increase quickly with the size, much more so than the number of modes.  

As the grain size increases, the number of skeletal vibration modes (involving the whole structure) increases more rapidly than the number of
 modes of functional groups (involving only a small number of atoms of adequate chemical nature, and with adequate bonds to other atoms). The UIBs
are mainly due to the latter, while the continuum (mainly in the farther IR) is due to the former. Thus, the fraction of energy deposited 
in the grain that goes into UIB emission progressively decreases as the size increases. This is observed up to the largest structures 
attainable by the usual simulations (of order of 10$^{3}$ atoms). \rm                                                    

Tailoring the composition and structure of the grains so that they carry the UIBs is outside the scope of this 
paper (it was treated at length by Papoular 2011). The validity of our conclusions do not depend on these properties,
 provided the grains bear enough carbon dangling bonds at their periphery, in steady state, so they can capture ambient H atoms.

\section{Acknowledgments}
I thank the reviewer, Dr C. Boersma, for several useful comments.

\end{document}